\documentclass[prl,twocolumn,amsmath,amssymb]{revtex4}
\usepackage{latexsym,amsmath,graphics,graphicx}

\begin{document}
\title{Non-collinear magnetism of Cr nanostructures on Fe$_{3ML}$/Cu(001):
first--principles and experimental investigations}

\author{Samir Lounis$^1$}\email{s.lounis@fz-juelich.de}
\author{Matthias Reif$^2$}
\author{Phivos Mavropoulos$^1$}
\author{Leif Glaser$^2$}
\author{Peter H.~Dederichs$^1$}
\author{Michael Martins$^2$}
\author{Stefan Bl\"ugel$^1$}
\author{Wilfried Wurth$^2$}

\affiliation{$^1$Institut f\"ur
Festk\"orperforschung, Forschungszentrum J\"ulich, D-52425 J\"ulich,
Germany\\
$^2$Universit\"at Hamburg, Institut f\"ur Experimentalphysik,
Luruper Chaussee 149, D-22761 Hamburg, Germany}

\begin{abstract}
A combined experimental, using X-ray Magnetic Circular Dichroism and
theoretical investigation, using full-potential Korringa-Kohn-Rostoker (KKR)
Green function method, is carried out to study the spin structure of
small magnetic Cr adatom--clusters on the surface of 3 monolayers of
$fcc$ Fe deposited on Cu(001). The exchange interaction between the
different Cr adatoms as well as between the Cr atoms and the Fe atoms
is of antiferromagnetic nature and of comparable magnitude, leading
due to frustration to complex non-collinear magnetic
configurations. The presence of non-collinear magnetic coupling
obtained by {\it ab initio} calculations is confirmed by the
experimental results.
\end{abstract}
\maketitle


The microscopic understanding of the magnetic properties of small
transition metal particles has recently attracted a lot of interest.
Small clusters show magnetic behavior which is distinctly different
from the respective bulk material; for instance, even non-magnetic
materials can show magnetism as small clusters \cite{cox93}. Spin and
orbital magnetic moments are usually significantly enhanced when the
average coordination number of the atoms is reduced in a cluster
\cite{Lau02}. Of particular interest is the tendency of small clusters
to favor non-collinear spin structures. In particular, for clusters on
surfaces the formation of non-collinear spin structures is observed to
relax possible spin frustration which results from a competition
between intra-cluster and cluster-substrate interactions. A prototype
of such a system are small Cr clusters on a ferromagnetic {\it fcc}
Fe/Cu(001) substrate. The latter one has already challenged
experimentalists and theoreticians for several years and is generally
accepted to be ferromagnetic (FM) up 4 Fe
monolayers\cite{thomassen,asada,moroni}. On the other hand Cr atoms tend to couple
antiferromagnetically to each other and lead in some 
cases\cite{bluegel,Bergman06,robles} to 
non-collinear magnetism.

The aim of this work is to determine the complex magnetism of Cr
nanoclusters on Fe/Cu(001) using state-of-the-art theoretical and
experimental techniques. Compared to bcc Fe, where adatoms can
approach each other at second-neighbor positions at the closest, an
fcc(001) Fe substrate allows for a close-packed cluster geometry with
first-neighbor positions between adatoms. This enhances the
intra-cluster exchange interactions. Furthermore, compared to a Ni
substrate, a Fe substrate has a much stronger exchange interaction
with the adatoms and ad-clusters. These properties provoke
magnetically frustrated states in antiferromagnetic ad-clusters (such
as Cr) and make the particular combination of cluster atoms and
substrate intriguing.

We first present the {\it ab-initio} description of this system
showing thus the importance of competing intra-cluster and
cluster-substrate exchange interaction leading to non-collinear spin
structures. The second step consists on depicting the measurements
obtained with X-ray Magnetic Circular Dichroism (XMCD) in order to be
compared with theory.

$-$ {\it Theory.} Our calculations are based on the Local
Density Approximation (LDA) of density functional theory with the
parametrization of Vosko {\it et al.}\cite{vosko} and use the
full-potential non-collinear KKR-Green function
method\cite{papanikolaou,lounis}. In the first step the surface Green
functions are determined by the screened KKR method \cite{SKKR} for
three monolayers of Fe (Fe$_{3ML}$) on Cu(001), using the LDA lattice
parameter of Cu ($\sim 3.51~{\AA}$). The obtained magnetic order of Fe
is ferromagnetic, with moments for the surface layer of $m_S = 2.66
\mu_B$ and the two subsurface layers ($m_{S-1} = 2.09 \mu_B$ and
$m_{S-2} = 2.18 \mu_B$). 
In the second step we calculate the Green function of the adatoms on
the surface, by using the surface Green function as a reference. The
resulting cluster of perturbed potentials includes the potentials of
the Cr adatoms and the perturbed potentials of several neighboring
shells.
Angular momenta up to $l_{\mathrm{max}} = 3$ were included in
the expansion of the Green functions. 
We consider all impurities at the unrelaxed hollow position in the
first vacuum layer. The orientations assigned to the spin moments of
the impurities are always relative to the orientation of the substrate
moment, which we take as the global frame~\cite{lounis}.

\begin{table}
\caption{Size and angles of magnetic moments in the calculated
clusters in the non-collinear configuration. The atoms separated by -- 
have the same magnetic moments and rotation angles.}
\begin{tabular}{lcrrr}
\hline
Cluster & Atom   & $\mathrm{M}\ (\mu_B)$ & $\theta\ (^\circ)$    & $\phi\ (^\circ)$ \\
\hline

Trimer & A, B--C & 2.57, 2.92  & 77, 156  & 180, 0\\
\hline
Tetra 1& A--D, B--C & 2.50,  2.50   & 111, 111  & 0, 180  \\
\hline
Tetra 2 &A, B--D, C     & 2.85, 2.87, 2.31  & 172, 176, 13  &  0, 0, 180 \\
\hline
Penta &A, B, C     & 2.44, 2.47, 2.17  & 138, 85, 46  &  0, 180, 180\\
         &D, E     & 2.48, 2.89 &  155, 164  &  0, 0 \\
\hline
\end{tabular}
\end{table}
We start with single Cr adatoms, which are antiferromagnetically (AF)
coupled to the substrate with an energy difference $\Delta E_{FM-AF} =
564.9$~meV to the ferromagnetic (FM) state, showing the strength of
the interaction with Fe. The preference of the AF configuration shows
up also in a considerably larger Cr local moment of $3.30 \mu_B$
compared to the metastable FM configuration ($2.80 \mu_B$). Having
established the strong AF coupling of the single adatom, we turn to
the adatom dimers. We considered the two adatoms placed as first
neighbors with different magnetic configurations. As it turns out the
parallel orientation within the dimer, with both atoms
antiferromagnetically oriented to the substrate, is the most stable
state with Cr moments of 3.02$\mu_B$.  Surprisingly, we did not find a
non-collinear solution which would be expected by the following
argument: Both Cr atoms tend to couple AF to the substrate as we have
seen for single adatoms, but they also tend to couple AF to each other
as expected for elements having a half filled $d$-band and as
explained by the Alexander-Anderson model\cite{anderson,lounis}. These
competing interactions, should lead to magnetic frustration and to a
non-collinear solution. However, when we introduce a rotation of the
moments in our {\it ab-initio} calculations, both impurities relax
back to the AF coupling to the substrate even if there is an AF
between the dimer atoms.  In order to explain this we introduce a
Heisenberg model with the exchange parameters calculated from a fit to
total energy calculations. We assume a classical spin Hamiltonian of
the form
${H} = - \frac{1}{2}\sum_{i \neq j}{J}_{ij}{\vec{e}_i \vec{e}_j}$, 
where $\vec{e}$ is a unit vector defining the direction of the
magnetic moment and $i$ and $j$ indicate the dimer atoms and their
first Fe neighbors. Taking into account only first-neighbor
interactions and neglecting the rotation of Fe moments, we rewrite the
Hamiltonian for the dimer in the form:
${H}  = - J_{\mathrm{Cr-Cr}} \cos(\theta_1 + \theta_2)
 - 4 J_{\mathrm{Cr-Fe}} (\cos\theta_1+\cos\theta_2)$. 
The angle defining the non-collinear solution is obtained after a
minimization of the Heisenberg Hamiltonian:
$\cos(\theta_1)=\cos(\theta_2)=-2J_{\mathrm{Cr-Fe}}/ J_{\mathrm{Cr-Cr}}$. 
If $2|{J_{\mathrm{Cr-Fe}}|>|J_{\mathrm{Cr-Cr}}}|$, the angle is not
defined and the non-collinear solution does not exist. This is
realized in the present case: $2|{J_{\mathrm{Cr-Fe}}}|=2\times 80.8
\mathrm{meV}
> |{J_{\mathrm{Cr-Cr}}}| = 77.6$~meV. Note that the Cr-Fe coupling
constants are considerably smaller than for the single adatom.
\begin{figure}
\begin{center}
\includegraphics*[width=\linewidth]{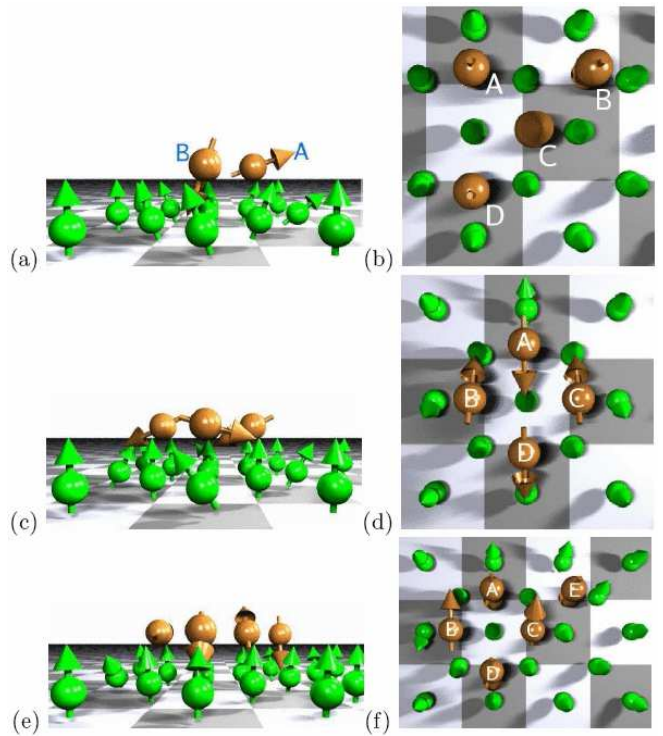}
\caption{(Color online) Complex magnetic states of Cr ad-clusters on Fe$_{3ML}$/Cu(001)
surface: (i) Side view of the trimer is shown in (a) with
two Cr atoms pointing
down (the second one cannot be seen), (ii) the front
view of the T-shape tetramer  (tetramer 2) is presented in (b),
(iii) both sides of tetramer 1 are shown in
(c) and (d), and finally the pentamer is shown in (e) and (f).}
\label{Cr-trimer-front}
\end{center}
\end{figure}

Following the same procedure as for the dimer we first investigated
several collinear magnetic configurations for the most compact trimer
on the surface, which has the shape of an isosceles rectangular
triangle (see Fig.~\ref{Cr-trimer-front}(a)). After self-consistency all
different collinear configurations converged to the FM solution with
drastically reduced moments, {\it i.e} 0.19$\mu_B$ for atom A and
0.42$\mu_B$ for both atoms B and C. If we allow the directions of the
moments to rotate, the moments strongly increase to ``normal'' values
(2.57$\mu_B$ for atom A and 2.92$\mu_B$ for atoms B and C). As seen
earlier, the AF interaction within the dimer was not high enough to
compete with the adatom-substrate interaction. Addition of a third
adatom to the system forces a rotation of the moments. The two second
neighboring impurities B and C have a moment tilted down by an angle
of 156$^\circ$ and adatom A moment is tilted up with an angle of
77$^\circ$ (Table 1). Thus due to the reduced Fe-Cr coupling the
trimer atoms show a nearly antiferromagnetic configuration. The total
energy differences between the non-collinear and collinear solution is
$\Delta E_{\mathrm{Ncol}-\mathrm{FM}} = -360$~meV/adatom.

We extended our study to bigger clusters, namely tetramers and
pentamer. Two types of tetramers were considered: tetramer 1 is the
most compact and forms a square (Fig.~\ref{Cr-trimer-front}(d)), while
tetramer 2 has a T-like shape (Fig.~\ref{Cr-trimer-front}(b)).
Concerning tetramer 1 (Fig.~\ref{Cr-trimer-front}(c) and (d)), two
collinear configurations were obtained: the AF solution, where all
impurities are coupled antiferromagnetically to the substrate, and a
Ferri solution where in each two neighbors within the tetramer are
oriented AF to each other.  The Ferri state is less stable than the AF
as can be seen from the energy difference $\Delta
E_{\mathrm{Ferri}-\mathrm{AF}} = 543.8$~meV/adatom. The Ferri
configuration is characterized by magnetic moments of $-2.45$~$\mu_B$
and 2.29~$\mu_B$ whereas in the AF configuration the atoms carry lower
moments (1.94~$\mu_B$). However, the ground state configuration is
non-collinear (Fig.~\ref{Cr-trimer-front}(c) and (d)) with an energy
difference $\Delta E_{\mathrm{AF}-\mathrm{Ncol}} = 80$~meV/adatom and
a magnetic moment of 2.5~$\mu_B$ carried by each impurity. One notices
that the first neighboring adatoms are almost AF coupled to each
other(the azimuthal angle $\phi$ is either equal to 0$^0$ or to
180$^0$) with all moments rotated by the angle $\theta$ = 111$^\circ$.

For tetramer 2 (see Fig.~\ref{Cr-trimer-front}(b)) we obtained several
collinear magnetic configurations. The most favorable one is
characterized by an AF coupling of the three corner atoms with the
substrate. The moment of adatom C, surrounded by the remaining Cr
impurities, is then forced to orient FM to the substrate.  When we
allow for the direction of the magnetic moment to relax, we get a
non-collinear solution having a similar picture, energetically close
to the collinear one ($\Delta E_{\mathrm{col}-\mathrm{Ncol}} =
2.3$~meV/adatom). Adatom C has now a moment somewhat tilted by
13$^\circ$ ($\mu=$~2.31~$\mu_B$) whereas adatom A has a moment tilted
in the opposite direction by 172$^\circ$
($\mu=$~2.85~$\mu_B$). Adatoms B and D have a moment of 2.87~$\mu_B$
with an angle of 176$^\circ$.  We note that tetramer 1 with a higher
number of first neighboring adatom bonds (four instead of three for
tetramer 2) is the most stable one ($\Delta
E_{\mathrm{tet2}-\mathrm{tet1}} = 14.5$~meV/adatom) with the
non-collinear solution shown in Fig.~\ref{Cr-trimer-front}(c) and (d).

To study the pentamer, we have chosen a structural configuration
(Fig.~\ref{Cr-trimer-front}(e) and (f)) with the highest number of
first neighboring adatom bonds (five). This pentamer consists on a
tetramer of type 1 plus an adatom (E) and is characterized by a
non-collinear ground state.  Let us understand the solution obtained
in this case: Tetramer 1 is characterized by a non-collinear almost
in-plane magnetic configuration (see Fig.~\ref{Cr-trimer-front}(e) and
(f)). As we have seen, a single adatom is strongly AF coupled to the
substrate. However, when one moves it closer to the tetramer it
affects primarily the first neighboring impurity (adatom C) by tilting
the magnetic moment from 111$^\circ$ to 46$^\circ$. Adatom E is also
affected by this perturbation and experiences a tilting of its moment
from 180$^\circ$ to 164$^\circ$. As a second effect, the second
neighboring adatom, A, is also affected and suffers a moment rotation
from 111$^\circ$ to 138$^\circ$.  The AF coupling between first
neighboring adatoms is always stable, thus adatom D has also a moment
rotated opposite to the magnetization direction of the substrate with
an angle of 155$^\circ$ ($\mu =$~ 2.48~$\mu_B$). As adatom B tends
to couple AF to its neighboring Cr adatoms, its magnetic moment tilts
into the positive direction with an angle of 85$^\circ$.

$-$ {\it Experiment.} The magnetic properties of the deposited clusters
have been determined experimentally using XMCD. The experiments were
performed at the beamline UE56/1- PGM at the BESSY II storage ring in
Berlin.  The mass selected chromium clusters were generated using a
UHV-cluster source \cite{Lau05} and deposited in-situ onto ultrathin
Fe layers epitaxially grown on a Cu(100) surface. The iron multilayers
were prepared by evaporating iron from a high purity iron sheet onto
the clean copper crystal. Subsequently the iron films in the thickness
range of $\sim$3-5 monolayers (ML) were magnetized perpendicular to
the surface plane using a small coil.  The magnetization of the iron
films was monitored by recording Fe 2p XMCD spectra. Before cluster
deposition Argon multilayers were frozen onto the iron surface at
temperatures below 30K. A layer thickness of ~10ML of argon was used
to ensure soft landing conditions \cite{Lau03}. After depositing the
clusters into the argon buffer layers, the remaining argon was
desorbed by flash heating the crystal to $\sim$80K. Low sample
temperatures in the range of 30K and low Cr coverages of 3\% of a
monolayer (ML) were used to prevent cluster-cluster interaction. The
measurements have been carried out at a base pressure $p < 3\cdot
10^{-10}$mbar.  Spectra have been measured from the cluster samples
prepared as described above, in a size range of 1 to 13 atoms per
cluster. Every step of the preparation has been checked using X-ray
Photoelectron Spectroscopy (XPS) and/or X-ray Absorption Spectroscopy
(XAS). The absorption signal has been measured using the Total
Electron Yield TEY, i.e. the sample current.

\begin{figure}
\begin{center}
\includegraphics*[width=1\linewidth]{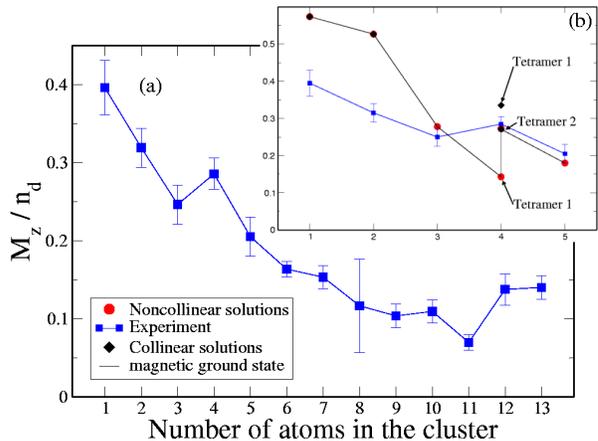}
\caption{(Color online) Spin moment per number of d-holes per atom versus cluster
size. Fig.a shows the experimental results for cluster-sizes up to
13 adatoms while Fig.b gives the comparison with {\it ab initio}
results for cluster-sizes up to 5 adatoms. Blue squares describe the
experimental values, black diamonds show represent collinear
configurations, red circles correspond to non-collinear
configurations and a black line connects the ratios obtained in the
magnetic ground states.} \label{seff-curve}
\end{center}
\end{figure}
\begin{figure}
\begin{center}
\includegraphics*[angle=270,width=0.8\linewidth]{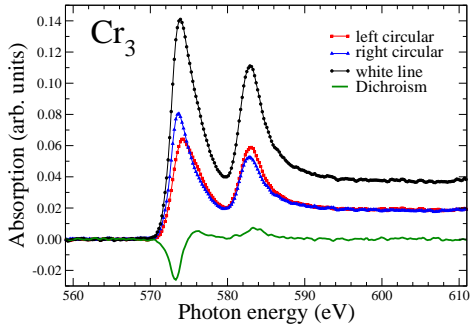}
\caption{(Color online) Absorption, difference and sum spectra of a
Cr$_3$ cluster.}
\label{spectra}
\end{center}
\end{figure}
The experimental results for the magnetic moments per d-hole shown
in Fig.\ref{seff-curve} have been determined applying XMCD sum rules
\cite{Laan99}. After a standard background treatment, difference and
sum spectra are generated by subtraction and addition of the x-ray
absorption spectra measured with different photon helicity. A
typical example for a Cr$_3$-cluster is shown in Fig.\ref{spectra}.
For the application of the sum rules, the areas of the 2p$_{3/2}$
and 2p$_{1/2}$ peaks of the difference spectra have been integrated.
The 2p$_{3/2}$ peak which shows positive and negative contributions
in the difference spectra has been integrated over both
contributions. The values for the spin and orbital magnetic moments
(not shown) per d-hole have been corrected for a degree of circular
polarization of 90\%. The contribution of the spin magnetic dipole
operator $\langle T_z\rangle$ in the values of the spin moments is
presently ignored.

We note that we have applied the XMCD sum rules without empirical
corrections which have been proposed for chromium. We are well aware
of the fact that due to the problems in the application of the XMCD
sum rules for the early transition metals with smaller spin-orbit
splitting the absolute values for the spin magnetic moments might be
too small by up to a factor of two. However, this will not affect the
relative trends seen in the experimental data
(Fig.~\ref{seff-curve}.(a)). One notices the strong decrease of
$S_{eff}/n_d$ with increasing cluster size which is due to the
appearance of antiferromagnetic or non-collinear structures as
calculated by theory. The qualitative and quantitative trends observed
in the experimental results agree very well with the theoretical
results (Fig.~\ref{seff-curve}.(b) for Cr-atoms to
Cr-pentamers. Although the theoretical values for Cr-atoms and dimers
lie somewhat higher than the experimental values (including error
bars) the agreement can still be judged to be very good in view of the
remaining experimental uncertainties discussed above.  Even details as
the increase of spin magnetic moment from trimer to tetramer can be
addressed. In order to understand the peak formed for the tetramer we
compare in Fig.~\ref{seff-curve}.(b) the ratio between the moment
along the $z$-direction (defined by the magnetization of the substrate)
and number of holes per atom obtained by theory for the different
geometrical and spin structures. Black circles show the ratio
calculated by taking into account the collinear solutions and the red
ones show the ratio calculated from the non-collinear solutions. The
black line connects the ratio obtained in the magnetic ground states.
The non-collinear tetramer 1 has a much lower value than what was seen
experimentally whereas the collinear tetramer 1 and tetramer 2 give a
better description of the kink seen experimentally. With regards to
the small energy difference ($\Delta E=14.5V$~meV) between the two
tetramers we considered, we believe that the experimental value can be
understood as resulting from an average of non-collinear tetramers 1,
collinear and non-collinear tetramer 2. We believe that this explains
why the tetramer ratio value is higher than the one obtained for a
trimer.  The trimer and the pentamer are clearly well described by the
theory and fit to the experimental measurements.

To summarize, the competing exchange interactions in Cr-clusters and
with Fe surface atoms affect strongly  the complex magnetic
structures leading thus to non-collinear magnetism in some cases.
This is predicted by DFT calculations and confirmed by XMCD
measurements. Considering the complexity of the different clusters
and the non-collinear spin structures and the difficult experimental
analysis the quantitative agreement between theory and experiment is
quite satisfying. A mechanism is proposed to explain the kink
observed experimentally for the tetramer in terms of geometrical
configuration average.  For further corroboration of this result
detailed STM investigations with spin-polarization analysis could be
envisioned.

This work was financed by the Priority Program ``Clusters in Contact
with Surfaces'' (SPP 1153) of the Deutsche Forschungsgemeinschaft.


\begin{thebibliography}{99}

\bibitem{cox93}{A. J. Cox, J. G. Louderback and L. A. Bloomfield,
Phys. Rev. Lett {\bf 71}, 923 (1993).}


\bibitem{Lau02}{J. T. Lau, A. F{\"o}hlisch, R. Nietubyc, M. Reif and W. Wurth, 
Phys. Rev. Lett. {\bf 89}, 57201 (2002).}

\bibitem{thomassen} J. Thomassen, F. May, B. Feldmann, M. Wuttig, and H. Ibach  
Phys. Rev. Lett. {\bf 69}, 3831 (1992).

\bibitem{asada} {T. Asada, S. Bl\"ugel, Phys. Rev. Lett. {\bf 79}, 507 (1997).}

\bibitem{bluegel} D. Wortmann, S. Heinze, Ph. Kurz, G. Bihlmayer, S. Bl\"ugel,
Phys. Rev. Lett. {\bf 86}, 4132 (2001).

\bibitem{Bergman06} A. Bergman, L. Nordstr\"om, A. B. Klautau, S. 
Frota-Pess\^{o}a, and O. Eriksson, Phys. Rev. B {\bf 73}, 174434 (2006).

\bibitem{robles} R. Robles, L. Nordstr\"om, accepted in Phys. Rev. B (2006).

\bibitem{vosko} {S. H. Vosko, L. Wilk, and M. Nusair,
J. Chem. Phys. {\bf 58}, 1200 (1980).}


\bibitem{papanikolaou} {N. Papanikolaou, R. Zeller, and P. H. Dederichs, 
J. Phys.: Condens. Matter.
{\bf 14}, 2799 (2002).}

\bibitem{SKKR}{K. Wildberger, R. Zeller, and P. H. Dederichs, 
Phys. Rev. B {\bf 55} 10074 (1997) and references
therein.}

\bibitem{stepanyuk} {V. S. Stepanyuk, 
W. Hergert, P. Rennert, B. Nonas, R. Zeller, and P. H. Dederichs, 
Phys. Rev. B {\bf 61}, 2356 (2000).}

\bibitem{moroni} {E. G. Moroni, G. Kresse and J. Hafner, 
J. Phys.: Condens. Matter {\bf 11}, L35 (1999).}



\bibitem{lounis}{S. Lounis, Ph. Mavropoulos, P. H. Dederichs, S. Bl\"ugel, 
Phys. Rev. B {\bf 72}, 224437 (2005).}

\bibitem{anderson}{S. Alexander and P. W. Anderson Phys. Rev. {\bf 133},
A1594 (1964).}

\bibitem{Lau05}{J. T. Lau {\it et al}, 
Rev. Sci. Instr. {\bf 76}, 063902 (2005).}

\bibitem{Lau03}{J. T. Lau, H.-U. Ehrke, A. Achleitner, and W.Wurth,
Low Temperature Physics, {\bf 29}, 223 (2003).}

\bibitem{Laan99}{G. van der Laan, J. Synchrotron Rad., {\bf 6}, 694 (1999).}

\end{thebibliography}
\end{document}